\documentclass[journal]{IEEEtran}
\usepackage{amsmath,amsfonts}
\usepackage{algorithmic}
\usepackage{algorithm}
\usepackage{array}
\usepackage[caption=false,font=normalsize,labelfont=sf,textfont=sf]{subfig}
\usepackage{authblk}
\usepackage{textcomp}
\usepackage{stfloats}
\usepackage{url}
\usepackage{verbatim}
\usepackage{graphicx}
\usepackage{cite}
\usepackage[numbers]{natbib}
\usepackage{multirow}
\usepackage{lineno,hyperref}
\usepackage{graphicx}
\usepackage{subcaption}
\usepackage{comment}
\usepackage[table]{xcolor}
\graphicspath{ {images/} }
\newcommand\new[1]{{\color{black}#1}} 

\begin{document}

\title{Assessing Vulnerability in Smart Contracts: The Role of Code Complexity Metrics in Security Analysis}

\author[1]{Masoud Jamshidiyan Tehrani}

\affil[1]{Universit\`a della Svizzera italiana, Lugano, Switzerland \\ masoud.jamshidiyantehrani@usi.ch}



\maketitle

\begin{abstract}
\new{Software built on poor structural patterns often shows higher exposure to security defects. When code differs from established best practices, verification and maintenance become increasingly difficult, thereby raising the risk of unintentional vulnerabilities. In the context of blockchain technology, where immutable smart contracts handle high-value transactions, the need for strict security assurance is important. This research analyzes the utility of software complexity metrics as diagnostic tools for identifying vulnerable Solidity smart contracts. We evaluate the hypothesis that complexity measures serve as vital, complementary signals for security assessment. Through an empirical examination of 21 distinct metrics, we analyzed their inter-dependencies, statistical association with vulnerabilities, and discriminative capabilities. Our findings indicate a significant degree of redundancy among certain metrics and a relatively low correlation between any single metric and the presence of vulnerabilities. However, the data demonstrates that these metrics possess strong power to distinguish between secure and vulnerable code when analyzed collectively. Notably, with only three exceptions, vulnerable contracts consistently exhibited higher mean complexity scores than their neutral counterparts. While our results show a statistical association, we emphasize that complexity is an indicator rather than a direct cause of vulnerability.}
\end{abstract}


\section{Introduction \& Motivation}
Ethereum~\cite{buterin2014next}, is a decentralized computing platform and operating system based on blockchain technology. It is an open-source platform that caters to diverse business scenarios by offering Solidity, a Turing-complete programming language, and the Ethereum Virtual Machine (EVM). Developers can utilize these tools to deploy smart contracts~\cite{szabo1996smart} on the Ethereum blockchain. smart contracts find widespread applications in various industries, including decentralized finance (DeFi) within financial services~\cite{bartoletti2017empirical}, infrastructure development~\cite{zhang2017distributed}, the internet of things (IoT)~\cite{christidis2016blockchains}, the gaming industry~\cite{bartoletti2017empirical}, healthcare~\cite{griggs2018healthcare, alhadhrami2017introducing}, the metaverse~\cite{jeon2022blockchain}, and numerous other sectors~\cite{bartoletti2017empirical}. In contrast to traditional applications, smart contracts are executed automatically in shortcodes, which safeguards them against manipulation during the execution process. Open-source availability is a common characteristic of widely-used smart contracts, infusing trust among users regarding the integrity of the code. Since smart contracts operate on distributed nodes within the blockchain, once a contract is executed and deployed in a block, it becomes quite challenging to upgrade or modify the contract~\cite{bui2021evaluating}. This is due to the inherent immutability of the blockchain, where tampering with any block is virtually impossible. Furthermore, smart contracts have the capability to hold funds within their associated blocks on the blockchain. These unique properties of smart contracts make them an attractive target for hackers. While the underlying blockchain technology itself is highly secure and resistant to hacking attempts, smart contracts can still be vulnerable to attacks. This vulnerability often stems from factors such as programmer negligence or errors in the code. Therefore, implementing robust vulnerability detection mechanisms becomes essential to identify and address any potential security flaws in smart contracts prior to their deployment~\cite{wang2021ethereum,atzei2017survey}. The security of smart contracts is of utmost importance, as evidenced by the significant amount of money lost in real-world smart contract vulnerabilities each month. A notable example is the attack on the Decentralized Autonomous Organization (DAO) in 2016, where 3.6 million Ether was stolen using a simple recursive vulnerability~\cite{mehar2019understanding}. Additionally, according to blockchaingroup.io \footnote{https://blockchaingroup.io/compliance-and-regulation/top-10-crypto-losses-of-2024-hacks-frauds-and-exploits/}, the top 10 crypto hacks in 2024 alone resulted in the theft of 1.03 billion US dollars by hackers. Table~\ref{tab:hacks} provides an overview of the amounts stolen in each smart contract hack. These staggering numbers serve as evidence of the critical importance of ensuring the security of smart contracts.\\
Complexity metrics are used in this study because they provide a quantifiable measure of a smart contract’s structural intricacy\cite{tonelli2023smart}, which has been historically linked to software defects and vulnerabilities~\cite{sultana2021using, agarwal2022cyclomatic, shin2011initial}. Their use is justified and motivated by several key reasons.

\textbf{Relevance to vulnerability prediction:}
Complexity has long been associated with software defects in traditional programming languages. Studies have shown that as complexity increases, the likelihood of errors and vulnerabilities also rises~\cite{shin2008complexity}. Since smart contracts execute autonomously and handle financial assets, even minor defects can lead to catastrophic failures. By using complexity metrics, the study aims to assess whether similar patterns hold for Solidity contracts.

\textbf{Objectivity and Quantifiability:}
Complexity metrics provide an objective and reproducible way to measure code characteristics~\cite{chowdhury2010can}. Unlike qualitative assessments (e.g., code readability or subjective security evaluations), these metrics can be systematically computed, analyzed, and compared across large datasets. This makes them a suitable choice for large-scale vulnerability detection efforts.

\textbf{Practical Integration into Automated Tools:}
Automated vulnerability detection tools rely on measurable features to assess risk~\cite{harer2018automated}. Complexity metrics are easy to extract from code and can be directly integrated into static analysis tools.

By focusing on complexity metrics, this study ensures a scalable, automated, and objective approach to understanding vulnerability trends in smart contracts. While other metrics may also provide valuable insights, complexity remains a fundamental property that directly influences the security and maintainability of smart contracts.

The rest of the paper is structured as follows. Section two presents our contribution and outlines the research questions we aim to answer. Section three provides background and related work, offering the necessary knowledge to understand the motivation of this study and reviewing previous work on the topic. Section four details each complexity metric used in this study. Section five describes the dataset utilized. Section six, the experimental setup, explains the techniques and analysis methods employed to answer the research questions. Section seven presents the results of our analysis, addressing each research question in detail and analyzing the answers. Section eight, the discussion, explores how this study's findings might influence smart contract best practices or developer guidelines and suggests future directions. Section nine, the threat to validity, explains the limitations and biases of this study and how they were addressed. Lastly, section ten concludes the paper, summarizing the answers to the research questions and the overall findings.

\section{Contribution \& Research Questions}
\begin{figure*}[ht!]
\centering
\includegraphics[width=1.0\textwidth]{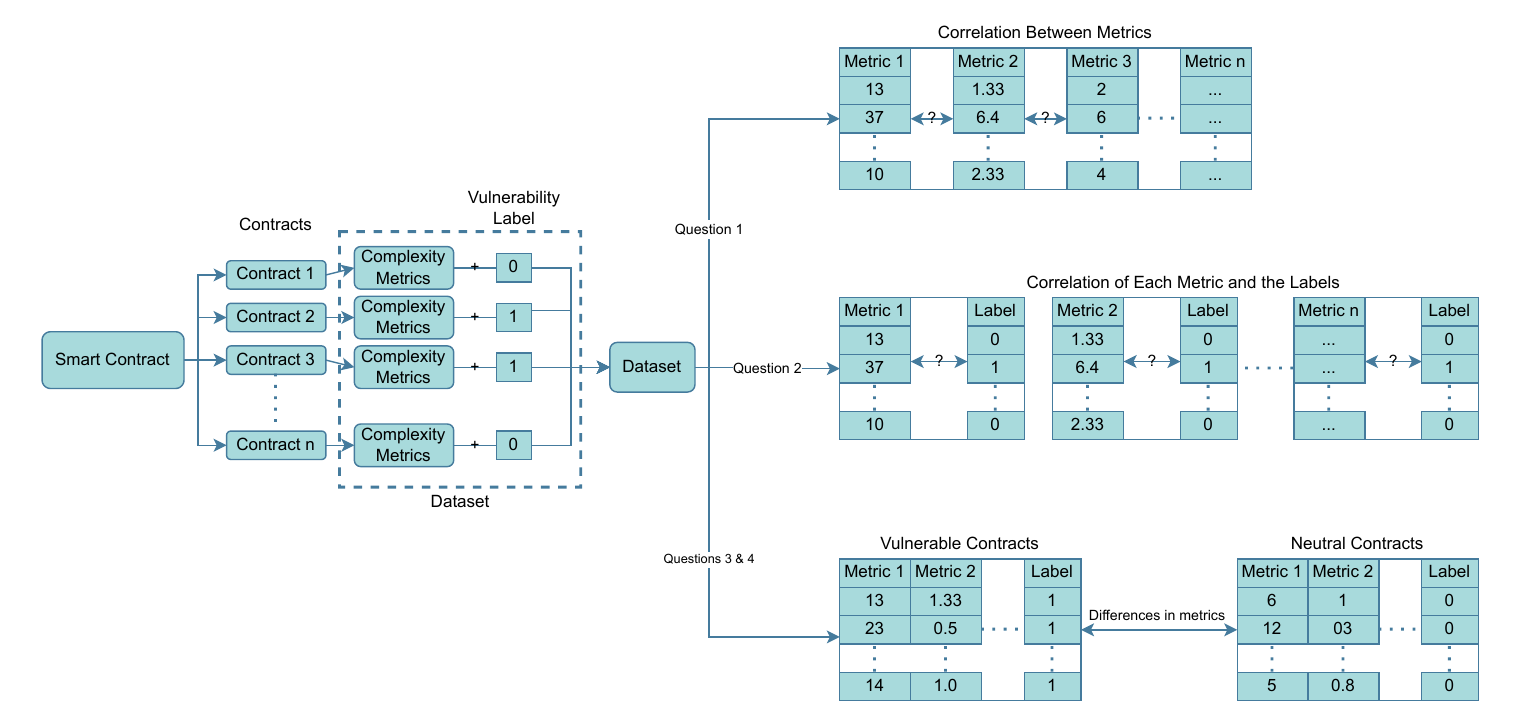}
\caption{Contribution and outline of this paper}
\label{fig:rmap}
\end{figure*}

The primary focus of this study is to investigate the impact of complexity metrics on vulnerabilities present in Solidity smart contracts. While existing static analysis tools are designed to detect specific vulnerability patterns\cite{kushwaha2022ethereum}, complexity metrics offer a complementary approach to identifying vulnerable areas within smart contracts. Our specific research objectives are as follows:
\begin{itemize}
    \item Determine whether there are noticeable differences in complexity metrics between vulnerable and non-vulnerable functions. If such differences exist, we aim to identify which complexity metrics are most effective in distinguishing between the two.
    \item Explore the feasibility of using complexity metrics to predict vulnerable contracts among all contracts within a smart contract ecosystem. 
\end{itemize}
This investigation aims to assess whether complexity metrics can serve as indicators for identifying vulnerable code locations. Furthermore, we aim to highlight the significance of complexity metrics as valuable complementary features for vulnerability assessment and provide insights into the individual power of each metric. Thus, we address four research questions related to software metrics and vulnerability detection:
\begin{itemize}
    \item \textbf{RQ1: Are there correlations among complexity metrics?} Our objective is to identify potential redundancies or inconsistencies among the various complexity metrics available.
    \item \textbf{RQ2: Is there a correlation between each complexity metric and the existence of vulnerability?} We aim to determine whether certain complexity metrics are associated with the presence of vulnerabilities in individual contracts.
    \item \textbf{RQ3: Are the metrics different in vulnerable and neutral codes?} We seek to investigate whether complexity metrics can effectively distinguish between functions with reported vulnerabilities and those without reported vulnerabilities.
    \item \textbf{RQ4: How different are the metrics in vulnerable codes compared to neutral ones?} \new{We examine the strength of the association between high complexity values and the presence of vulnerabilities to determine their utility as predictive indicators.} This approach enables a more targeted and efficient allocation of resources toward securing smart contracts.
\end{itemize}
The graphical representation of this paper's contributions and a brief outline of how the research questions are addressed can be found in Figure~\ref{fig:rmap}.

\section{Background \& Related Work}
Researchers have been actively proposing various methods and tools to identify security vulnerabilities within software.

\subsection{Static Analysis}
Static code analysis involves examining the source code of a smart contract without executing it, aiming to detect potential vulnerabilities through code inspection. Slither~\cite{feist2019slither} is a Python 3-based open-source static analysis framework that relies on the Solidity compiler (Solc) as its dependency. It offers swift, reliable, precise, and comprehensive information about smart contracts. Slither makes use of SlithIR, an intermediate representation designed for practical analysis of Solidity code. In addition to bug detection, it provides suggestions for code optimizations and enhances the comprehension of smart contract code. In addition to Slither, there are several other noteworthy static analysis tools for smart contracts, such as SolAnalyser~\cite{akca2019solanalyser} and EtherTrust~\cite{grishchenko2018ethertrust}. 

\subsection{Complexity and Vulnerability}
Defining program complexity is a crucial step in understanding software systems. According to Basili~\cite{basili1980qualitative}, complexity can be seen as a measure of the resources utilized by a system while interacting with a piece of software to accomplish a specific task. The interpretation of complexity depends on the interacting system, whether it is a computer or a programmer. In the context of a computer, complexity is determined by the execution time and storage requirements for computation. On the other hand, when considering the interaction between a program and a programmer, complexity is characterized by the challenges involved in tasks such as coding, debugging, testing, and modifying the software. The term ``software complexity'' often refers to the interaction between a program and a programmer engaged in programming activities~\cite{kearney1986software}.
\\
Security experts assert that complexity poses a significant threat to security~\cite{schneier2003beyond,mcgraw2006software}. Complex code can introduce subtle vulnerabilities that are challenging to detect and diagnose~\cite{mcgraw2006software}, thereby offering attackers greater opportunities for exploitation. Furthermore, complex code presents difficulties in comprehension, maintenance, and testing. Consequently, complex code is more prone to faults compared to simpler code. As attackers exploit these faults in a program, complex code becomes more vulnerable than its simpler counterparts. Although software complexity metrics in smart contracts have been widely explored, their precise relationship to vulnerabilities remains an open problem~\cite{agarwal2022cyclomatic, tonelli2023smart, vivar2020analysis}. However, the effectiveness of complexity metrics in identifying vulnerable code in traditional programs has been extensively studied and validated in various applications and programming languages~\cite{moshtari2013using, almogahed2022software}. Researchers have leveraged techniques like machine learning and statistical models to demonstrate the correlation between code complexity and the presence of vulnerabilities.
\\
In a study conducted by Shin and Williams~\cite{shin2008complexity}, a statistical analysis was performed to examine the relationship between complexity and software security. To assess the relationships between the metrics and vulnerabilities, the researchers employed two correlation coefficients: Pearson correlation coefficient (r) and Spearman rank correlation coefficient ($\rho$). Pearson correlation coefficient assumes a normal distribution of data and measures the strength of linear relationships between variables. On the other hand, the Spearman rank correlation coefficient is a non-parametric test that does not make any assumptions about the distribution of the data and measures the strength of monotonic relationships between variables.  Several complexity metrics, including (McCC) McCabe's cyclomatic complexity~\cite{mccabe1976complexity}, Strict cyclomatic complexity, (Nesting) the deepest level of nested control constructs such as ``if, while, for, switch, etc.'',(SLOC) the number of source code lines excluding comments, and (Stmt\_exe) the number of executable statements, were employed. The results of the analysis revealed the impact of complexity on software security. The study provides empirical evidence that supports the notion that complexity can indeed be considered an enemy of software security.
\\
In a study conducted by~\cite{alves2016software}, the researchers investigated the correlation between software metrics and the presence of security vulnerabilities in C and C++ programs. The analysis revealed that a significant number of the studied metrics exhibited the ability to distinguish between vulnerable and non-vulnerable codes. Some findings aligned with expectations, while others were unexpected. For instance, the vulnerable code was found to contain more comments than the code without vulnerabilities. Additionally, the study highlighted the presence of strong correlations between certain metrics.
\\
Two studies by Chowdhury et al.~\cite{chowdhury2010can,chowdhury2011using}, have identified empirical relationships between software metrics, including cohesion, coupling, complexity, and vulnerability using Spearman rank correlation. These studies categorized the metrics into two groups: design-level metrics and code-level metrics. Design-level metrics can be computed during the design phase, prior to the coding stage. On the other hand, code-level metrics are computed after the coding phase is completed.
\begin{itemize}
    \item Code-level complexity metrics include McCabe's cyclomatic complexity, Nesting, and SLOC. Strict cyclomatic Complexity is another metric that was measured which is identical to cyclomatic complexity except that the AND and OR logical operators are also counted as 1. Comment Ratio is also counted as a complexity measure which is the ratio of the number of comment lines to the number of code lines.
    \item Design-level complexity metrics encompass WMC (Weighted Methods per Class), DIT (Depth of Inheritance Tree), and NOC (Number of Children). WMC calculates the number of local methods defined within a class, DIT determines the maximum depth of a class in the inheritance hierarchy, and NOC represents the number of immediate sub-classes or derived classes. CBC is a metric that counts  the number of base classes (ancestors)
    \item Code-level coupling metrics consist of FanIn and FanOut. FanIn measures the number of inputs used by a function, while FanOut quantifies the number of outputs that are set by a function.
    \item  Design-level coupling metrics include DIT, NOC, CBC, RFC, and CBO. DIT and NOC are also considered complexity metrics. CBO indicates the level of coupling between object classes. More children in a class mean that it can be coupled to more methods and instance variables. Therefore, NOC and CBC measure coupling as well. RFC indicated the number of methods in the set, including inherited methods.
    \item Design-level cohesion metric refers to LOCM (Lack of Cohesion of Methods). LOCM evaluates the cohesion of methods within a class.
\end{itemize}
The findings of these studies indicate that complexity and coupling metrics in classic software, exhibit a positive correlation with the number of vulnerabilities. This implies that as the complexity and coupling of a software system increase, the likelihood of vulnerabilities also tends to increase. Conversely, cohesion demonstrated a negative correlation with vulnerabilities. This suggests that higher levels of cohesion within a software system are associated with a reduced number of vulnerabilities.
\\
In their research, Shin, Meneely, Williams, and Osborne~\cite{shin2010evaluating} conducted two empirical studies on two open-source projects to investigate the predictive power of three software metrics (code churn, complexity, developer activity) in identifying vulnerable locations. In their study, the researchers employed logistic regression along with four other classification techniques: decision tree, Random Forest, Naive Bayes, and Bayesian network. The findings revealed that out of the 28 metrics analyzed for both projects, 24 metrics successfully predicted vulnerabilities. Among these 24 metrics, 13 complexity metrics proved to be particularly useful. Examples of these complexity metrics include:
\begin{itemize}
    \item CountLineCode: Represents the total number of lines of code in a file.
    \item CountDeclFunction: Indicates the number of functions defined in a file.
    \item SumCyclomaticStrict: Calculates the sum of the number of conditional statements in a function.
    \item SumFanOut: Measures the sum of the number of assignments to parameters when calling a function or modifying global variables.
\end{itemize}
These complexity metrics played an important role in identifying potential vulnerabilities within the studied projects.

\subsection{Smart Contract Architecture}
\begin{figure}[h!]
\centering
\includegraphics[width=8cm]{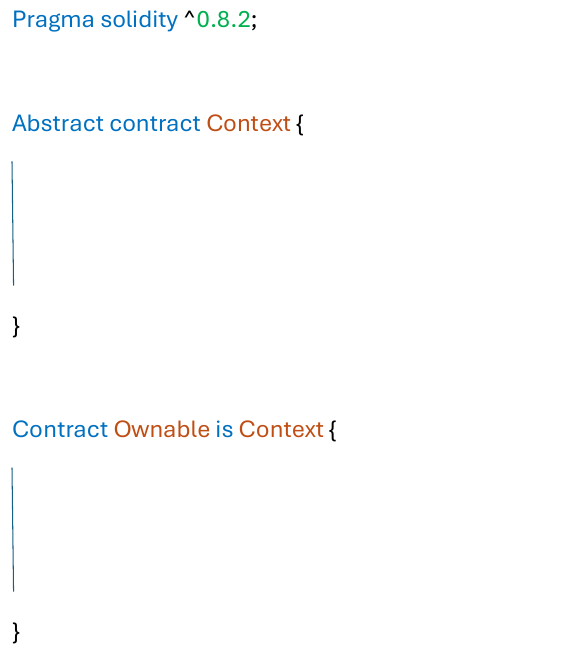}
\caption{General structure of a Solidity smart contract}
\label{fig:contracts}
\end{figure}

Solidity smart contract structures and vulnerabilities exhibit distinct characteristics compared to traditional programming languages.~\cite{buterin2014next}
\begin{itemize}
    \item Solidity has a reduced set of data types and libraries compared to most traditional programming languages. It is designed to be simple and deterministic for security and safety reasons.
    \item Solidity and other blockchain-specific languages have specific security concerns, such as vulnerabilities related to smart contract bugs, reentrancy, and transaction ordering. Traditional languages also have security concerns, but they differ based on their application domains and execution environments.
    \item  Solidity programming takes into account the concept of ``gas,'' which is a measure of the computational resources required to execute a smart contract. Each operation in Solidity consumes gas, and users must pay for the gas when executing a contract. Traditional languages don't have this gas mechanism as they are not executed on blockchains, and the cost of executing code is typically borne by the server or user's computing resources.
    \item In Solidity, state variables represent the persistent state of a smart contract on the blockchain. They are declared at the contract level and persist across function calls. Traditional languages may have variables with different scopes and lifetimes.
    \item Solidity allows the use of function modifiers, which are like reusable code snippets that can be applied to multiple functions in a contract to modify their behavior.
    \item In Solidity, functions can be marked as ``payable'', indicating that they can receive and handle cryptocurrency payments. Additionally, Solidity has a concept of gas usage for each function, which determines the cost of executing that function on the blockchain.
    \item Lastly, Solidity contains new data types such as ``address'', ``Bytes32'', and ``enum'', which further verify that software metrics in Solidity are different.
\end{itemize}
These distinctive properties of Solidity set it apart from other programming languages, encouraging us to revisit the use of complexity metrics. Although these metrics have been employed in previous studies (mentioned in section 3), the unique nature of smart contracts and the difficulties involved in the Solidity code demand a fresh evaluation of their effectiveness.
\\
In the context of smart contracts, multiple contracts can exist within a single smart contract, resembling the concept of objects in object-oriented programming languages. The diagram depicted in Figure~\ref{fig:contracts} provides an illustration of the structure of a Solidity smart contract.
Our objective is to conduct a comprehensive analysis of each contract individually and identify vulnerable contracts within the smart contract ecosystem. By focusing on each contract separately, we can gain insights into their specific vulnerabilities and assess their overall security. This approach allows us to pinpoint and address vulnerabilities at the contract level, enabling more targeted and effective security measures.

\begin{table}
\caption{Top 10 smart contract hacks and losses in 2024, according to blockchaingroup.io}
\centering
\label{tab:hacks}
\begin{tabular}{ll}
\hline
\textbf{smart contract}                                                               & \textbf{Loss}(\$)      \\
UwU Lend                                                                     & 19.3M          \\
Indodax                                                                      & 22M            \\
BinX Exchange                                                                & 26M            \\
Penpie                                                                       & 27M            \\
Hedgey Finance                                                               & 44M            \\
BtcTurk Hot Wallet                                                           & 55M            \\
DMM Bitcoin Private Key                                                      & 125M           \\
Orbit Chain Bridge                                                           & 197M           \\
WazirX                                                                       & 230M           \\
PlayDapp                                                                     & 290M           \\
\textbf{Total} & \textbf{1.03B} \\ \hline
\end{tabular}
\end{table}
\begin{table*}[ht]
\centering
\caption{Solidity complexity features analyzed by Solmet}
\label{tab:solmet}
\begin{tabular}{|l|l|}
\hline
\rowcolor[HTML]{EFEFEF} 
\textbf{SLOC}        & number of source code lines                                                        \\ \hline
\textbf{LLOC}        & number of logical code lines (lines without empty and comment lines)               \\ \hline
\rowcolor[HTML]{EFEFEF} 
\textbf{CLOC}        & number of comment lines                                                            \\ \hline
\textbf{NF}          & number of functions                                                                \\ \hline
\rowcolor[HTML]{EFEFEF} 
\textbf{WMC}         & weighted sum of McCabe's style complexity over the functions of a contract         \\ \hline
\textbf{NL}          & the deepest nesting level of control structures in functions summed for a contract \\ \hline
\rowcolor[HTML]{EFEFEF} 
\textbf{NLE}         & nesting level else-if                                                              \\ \hline
\textbf{NUMPAR}      & number of parameters                                                               \\ \hline
\rowcolor[HTML]{EFEFEF} 
\textbf{NOS}         & number of statements                                                               \\ \hline
\textbf{DIT}         & depth of inheritance tree                                                          \\ \hline
\rowcolor[HTML]{EFEFEF} 
\textbf{NOA}         & number of ancestors                                                                \\ \hline
\textbf{NOD}         & number of descendants                                                              \\ \hline
\rowcolor[HTML]{EFEFEF} 
\textbf{CBO}         & coupling between object classes                                                    \\ \hline
\textbf{NA}          & number of attributes (i.e. states)                                                 \\ \hline
\rowcolor[HTML]{EFEFEF} 
\textbf{NOI}         & number of outgoing invocations (i.e. fan-out)                                      \\ \hline
\textbf{Avg. McCC}   & McCabe's cyclomatic complexity                                                     \\ \hline
\rowcolor[HTML]{EFEFEF} 
\textbf{Avg. NL}     & Average number of NL                                                               \\ \hline
\textbf{Avg. NLE}    & Average number of nesting level else-if                                            \\ \hline
\rowcolor[HTML]{EFEFEF} 
\textbf{Avg. NUMPAR} & Average number of parameters                                                       \\ \hline
\textbf{Avg. NOS}    & Average number of statements                                                       \\ \hline
\rowcolor[HTML]{EFEFEF} 
\textbf{Avg. NOI}    & Average number of outgoing invocations                                             \\ \hline
\end{tabular}
\end{table*}
\section{Complexity Metrics}
The complexity metrics we employ in our analysis are obtained through the use of Solmet~\cite{hegedHus2019towards}, a static analysis tool specifically designed for extracting complexity metrics from Solidity code. Solmet enables us to extract a comprehensive set of 21 complexity metrics for each contract individually. Table~\ref{tab:solmet} provides a detailed overview of all the metrics extracted by Solmet.
\begin{itemize}
    \item \textbf{SLOC:} The Source Lines of Code (SLOC) metric captures the total number of lines in a software artifact, including comments and white spaces. This metric was measured in~\cite{shin2010evaluating} as CountLineCode.
    \item \textbf{LLOC:} Logical Lines of Code (LLOC) is a complexity metric that measures the number of executable lines of source code in a software artifact (lines without comments and white spaces). While LLOC primarily indicates the size of the codebase, prior research has shown a strong correlation between SLOC and complexity. This metric was named SLOC in other papers~\cite{shin2008complexity,chowdhury2010can,chowdhury2011using}
    \item \textbf{CLOC:} The number of comment lines (CLOC) metric, as utilized in prior studies~\cite{chowdhury2010can,chowdhury2011using}, plays a role in code-level complexity metrics by representing the comment ratio in software code. The comment ratio is calculated by dividing the number of comment lines by the number of code lines. By considering both CLOC and the number of lines of code (LLOC), the comment ratio can be determined, providing insights into the proportion of comments relative to the actual code. According to the findings in~\cite{malik2008understanding}, it is suggested that highly complex program units tend to have a higher comment ratio, indicating a larger number of comments associated with each line of code. This suggests that the comment ratio may provide some insights into the complexity of the code.
    \item \textbf{NF:} The number of functions (NF) metric was evaluated and demonstrated its usefulness in the study conducted by Shin et al. ~\cite{shin2010evaluating}. This metric focuses on quantifying the total number of functions within a contract. This was called RFC in~\cite{chowdhury2010can,chowdhury2011using}. The weighted version of this metric was also used in their work as WMC.
    \item \textbf{WMC} The weighted sum of McCabe's style complexity (WMC) metric, which considers the complexity of individual functions within a contract, has been discussed in previous studies such as Shin et al.~\cite{shin2008complexity,shin2010evaluating}, Chowdhury et al.~\cite{chowdhury2010can,chowdhury2011using}. WMC is an extension of the strict cyclomatic complexity metric that incorporates weights assigned to each function.
    \item \textbf{NL:} NL was also measured in~\cite{shin2008complexity,chowdhury2010can,chowdhury2011using} as Nesting, is the deepest nesting level of control structures in functions summed for a contract. The presence of highly nested control structures (such as if, while, for, switch, etc.) in a function can contribute to the complexity of a program entity. When control structures are deeply nested, it can make the code more intricate and challenging for programmers to comprehend.
    \item \textbf{NLE:} The Nesting Level of Else-If (NLE) metric specifically focuses on the nesting level of ``else-if'' control structures within the code. It is similar to the Nesting Level (NL) metric, but it is only counted for ``if'' and ``else-if'' control structures.
    \item \textbf{NUMPAR:} The NUMPAR metric, also known as the Number of Parameters, refers to the count of parameters or inputs that functions within a contract have.
    \item \textbf{NOS:} The number of statements (NOS) was also referenced as stmt\_exe in the study conducted by Shin et al.~\cite{shin2008complexity}. NOS counts conditional and executable statements in a contract. 
    \item \textbf{DIT:} The depth of inheritance tree (DIT) was also discussed as a design-level complexity and coupling metric in the paper referenced~\cite{chowdhury2010can,chowdhury2011using}. The maximum depth of a class in the inheritance tree, also known as the Depth of Inheritance Tree (DIT), refers to how deep a class is positioned within the hierarchy of inherited classes. A higher DIT value indicates that the class inherits a greater number of methods, which can contribute to increased complexity in predicting its behavior. The deeper a class is in the inheritance hierarchy, the more complex it becomes to understand and anticipate its functionality~\cite{chidamber1994metrics}.
    \item \textbf{NOA \& NOD:} In the research papers by Chowdhury et al.\cite{chowdhury2010can,chowdhury2011using}, the metrics ``number of ancestors'' (NOA) and ``number of descendants'' (NOD) were used and referred to as CBC and NOC, respectively. These metrics are related to the inheritance relationships among classes in an object-oriented system.
    \item \textbf{CBO:} The Coupling Between Object classes (CBO) is a metric that measures the number of other classes coupled to a specific class (or contract). It indicates the level of dependency and interaction between classes within an object-oriented system. A higher value of CBO suggests that a class is more tightly coupled to other classes, indicating a potentially higher complexity and interdependencies. Chowdhury et al.~\cite{chowdhury2010can,chowdhury2011using} conducted research where they measured the CBO metric and established its relationship with vulnerabilities.
    \item \textbf{NA:} The metric NA, which stands for Number of Attributes, represents the count of attributes associated with a class or a contract in Solidity. Attributes, in the context of Solidity contracts, refer to various modifiers or keywords that are added before a variable name to define its characteristics or structure. Examples of attributes in Solidity contracts include ``external,'' ``indexed'', ``internal,'' ``payable'' and others. This metric is counting the vocabulary size of a contract.
    \item \textbf{NOI:} The metric NOI (Number of Outgoing Invocations), also referred to as fan-out, quantifies the count of outgoing invocations or function calls from a specific function or method. It represents the number of other functions or methods that are called within the target function. The fan-out metric measures the degree of coupling or dependencies between the target function and other functions in the codebase. It indicates how extensively the target function relies on and interacts with other functions. A higher fan-out value suggests that the target function invokes or depends on a larger number of external functions. Fan-out was also previously measured in~\cite{shin2010evaluating, chowdhury2010can,chowdhury2011using}.
    \item \textbf{Avg. McCC:} The metric Avg. McCC (Average McCabe's Cyclomatic Complexity) is a measure of the average cyclomatic complexity of a software artifact, such as a function, module, or class. It quantifies the complexity of the control flow within the artifact based on the number of decision points and possible paths through the code. McCabe's cyclomatic complexity is calculated by counting the number of independent paths through the code, which is determined by the number of decision points, such as conditionals (if statements, switch cases), loops (for, while, do-while), and logical operators. The higher the cyclomatic complexity, the more complex and intricate the control flow of the code. The  McCC metric has been mentioned in the previously cited papers as a relevant measure of complexity and vulnerability.
    \item \textbf{Avg. :} Each of the metrics, namely Avg. NL (Average Nesting Level), Avg. NLE (Average Nesting Level of Else-If), Avg. NUMPAR (Average Number of Parameters), Avg. NOS (Average Number of Statements), and Avg. NOI (Average Number of Outgoing Invocations), calculates the sum of its respective metric for each contract. This sum is then divided by the number of functions present in the contract to obtain the average value.
\end{itemize}
Each metric that counts a number, such as the number of lines of code or the number of parameters, represents either the vocabulary size or the code size of a contract. These metrics have been highlighted in a prior study conducted by Peitek et al.~\cite{peitek2021program} as factors that require more attention and mental effort from programmers. As a result, they can increase cognitive load during code comprehension and impose a greater burden on the working memory of programmers. Consequently, these metrics can serve as indicators of potential vulnerabilities in a contract. By examining these metrics, we can gain insights into the complexity and potential risks associated with the code.
\begin{table}[ht]
\centering
\caption{Vulnerabilities labeled in the dataset}
\label{tab:labels}
\begin{tabular}{l}
\hline
\multicolumn{1}{c}{\textbf{Vulnerability Type}} \\ \hline
\rowcolor[HTML]{EFEFEF} 
Integer Underflow/Overflow                                       \\
Dangerous Ether Strict Equality \\
\rowcolor[HTML]{EFEFEF} 
Reentrancy                                      \\
Timestamp Dependency                                             \\
\rowcolor[HTML]{EFEFEF} 
Block Number Dependency                                 \\
Dangerous Delegatecall                                      \\
\rowcolor[HTML]{EFEFEF} 
Ether Frozen                                     \\
Unchecked External Call                                        \\
\end{tabular}
\end{table}
\section{Dataset and Labeling}

For our research, we utilized the dataset provided by Liu et al~\cite{liu2023rethinking}. This paper introduces IR-Fuzz, an automated fuzzing framework designed to detect vulnerabilities in smart contracts through invocation ordering and crucial branch revisiting.

The dataset in the paper was obtained by crawling verified contracts from Etherscan\footnote{https://etherscan.io}, which are real-world smart contracts deployed on the Ethereum Mainnet. During this process, 5,074 duplicate contracts were removed by comparing the hash of the contract binary code. The final dataset consists of 12,515 smart contracts with available source code.  

As listed in Table~\ref{tab:labels}, eight types of vulnerabilities were focused on in the dataset: timestamp dependency (TP), block number dependency (BN), dangerous delegatecall (DG), Ether frozen (EF), unchecked external call (UC), reentrancy (RE), integer overflow (OF), and dangerous Ether strict equality (SE). For experimentation, all smart contracts from the dataset were deployed onto a local Ethereum test network.  

For ground truth labeling, vulnerability-specific patterns were defined for each type of vulnerability to provide an initial label, followed by manual verification to confirm whether a smart contract actually contained a given vulnerability. By using these predefined patterns (e.g., keyword matching), potentially vulnerable contracts were efficiently identified while reducing the time spent labeling safe contracts.

This paper uses the publicly available ``Resource 3'' of the dataset on their GitHub\footnote{https://github.com/Messi-Q/Smart-Contract-Dataset}. This ``Resource 3'' dataset includes 2,953 smart contracts and a total of 16,239 individual contracts which contained 258 vulnerable and 15981 neutral contracts.

\section{Experimental Setup}
To answer our questions and make them comparable to previous works, we employ several statistical techniques chosen based on the previous works on complexity metrics and the dataset.

\textbf{Spearman's Correlation Coefficient:} The Spearman correlation coefficient is a statistical measure that allows us to assess the monotonic relevance of each metric separately to the presence of vulnerabilities in smart contracts~\cite{hinkle2003applied}. It determines the strength and direction of the monotonic relationship between two variables and is a non-parametric method, which means it does not assume a specific distribution for the variables. This statistical measure is particularly useful when analyzing data without relying on the magnitudes of the values. Instead, it focuses on the ranks or order of the values. This property makes it robust and less sensitive to outliers in the data~\cite{cohen2013statistical}. By considering the ranks, the Spearman rank correlation can capture non-linear relationships between variables, making it a suitable choice for assessing the association between complexity metrics and vulnerability indicators~\cite{chowdhury2010can}.
The formula for calculating the Spearman correlation coefficient ($\rho$) is:
\begin{equation}
    \rho =\frac{1 - (6 * sum)}{(n * (n^{2} - 1))}
\end{equation}
In this formula, 'n' represents the number of observations, and 'sum' represents the sum of squared differences. The formula calculates the sum of squared differences between the ranks of corresponding observations for the two variables. It then divides this sum by a scaling factor to obtain the Spearman correlation coefficient ($\rho$). The coefficient ranges between -1 and 1, where values close to 1 indicate a strong positive monotonic relationship, values close to -1 indicate a strong negative monotonic relationship and values close to 0 indicate a weak or no monotonic relationship.
\\
In statistical analysis, the significance of a correlation is commonly evaluated using a p-value, which represents the probability of observing a correlation by chance. A smaller p-value indicates a higher level of confidence in the significance of the correlation. The threshold commonly used to determine statistical significance is 0.05, which corresponds to a 95\% confidence level. If the p-value is less than or equal to 0.05, it is considered statistically significant, indicating that the observed correlation is unlikely to occur by chance~\cite{cohen2013statistical}

\new{\textbf{Mann-Whitney U Test:} Since software complexity metrics often follow a non-normal distribution, we employed the Mann-Whitney U test~\cite{mann1947test} (also known as the Wilcoxon rank-sum test). This non-parametric test allows us to assess whether the distribution of metric values in vulnerable contracts differs significantly from that of neutral contracts, without assuming a normal distribution or equal sample sizes.}

In this test, if the p-value is smaller than the chosen significance level (in our case 0.05), you reject the null hypothesis and conclude that there is a significant difference between the \new{distributions} of the two \new{independent} groups~\cite{moore2007basic}.

\new{To quantify the magnitude of the differences between vulnerable and neutral contracts, we calculated Cliff's Delta ($\delta$)~\cite{MeisselYao2024CliffsDelta}. This non-parametric effect size measure estimates the probability that a value from the vulnerable group is strictly greater than a value from the neutral group, minus the reverse probability. We interpret the effect sizes using the thresholds defined by Romano et al.~\cite{romano2006appr}: negligible ($|\delta| < 0.147$), small ($0.147 \le |\delta| < 0.33$), medium ($0.33 \le |\delta| < 0.474$), and large ($|\delta| \ge 0.474$).}

\textbf{Confidence Interval:} The primary objective of computing a statistic from a random sample is to approximate the mean of the entire population. However, the accuracy of this approximation is always a concern. To address this issue, a confidence interval is employed, which provides a range of values likely to contain the true population parameter of interest. The confidence level for the interval is chosen by the specialist, typically set at 95\%. This means that if the population is sampled multiple times, the resulting interval estimates will encompass the true population parameter in approximately 95\% of cases~\cite{moore2007basic}. The confidence interval provides a valuable tool to gauge the reliability and precision of the sample statistic in estimating the characteristics of the entire population.
\begin{table}[ht]
\centering
\caption{Correlations between complexity metrics and the existence of Vulnerabilities}
\label{tab:spearman}
\begin{tabular}{|l|c|c|}
\hline
\rowcolor[HTML]{9B9B9B} 
\multicolumn{1}{|c|}{\cellcolor[HTML]{9B9B9B}\textbf{Metrics}} & \multicolumn{1}{l|}{\cellcolor[HTML]{9B9B9B}\textbf{Correlation Coefficient}} & \textbf{P-Value} \\ \hline
SLOC                                                           & 0.153                                                                         & \textless 0.05   \\ \hline
\rowcolor[HTML]{EFEFEF} 
LLOC                                                           & 0.132                                                                         & \textless 0.05   \\ \hline
CLOC                                                           & -0.077                                                                        & \textless 0.05   \\ \hline
\rowcolor[HTML]{EFEFEF} 
NF                                                             & 0.123                                                                         & \textless 0.05   \\ \hline
WMC                                                            & 0.130                                                                         & \textless 0.05   \\ \hline
\rowcolor[HTML]{EFEFEF} 
NL                                                             & 0.146                                                                         & \textless 0.05   \\ \hline
NLE                                                            & 0.144                                                                         & \textless 0.05   \\ \hline
\rowcolor[HTML]{EFEFEF} 
NUMPAR                                                         & 0.082                                                                         & \textless 0.05   \\ \hline
NOS                                                            & 0.160                                                                         & \textless 0.05   \\ \hline
\rowcolor[HTML]{EFEFEF} 
DIT                                                            & 0.099                                                                         & \textless 0.05   \\ \hline
NOA                                                            & 0.102                                                                         & \textless 0.05   \\ \hline
\rowcolor[HTML]{EFEFEF} 
NOD                                                            & -0.042                                                                        & \textless 0.05   \\ \hline
CBO                                                            & 0.182                                                                         & \textless 0.05   \\ \hline
\rowcolor[HTML]{EFEFEF} 
NA                                                             & 0.170                                                                         & \textless 0.05   \\ \hline
NOI                                                            & 0.165                                                                         & \textless 0.05   \\ \hline
\rowcolor[HTML]{EFEFEF} 
Avg. McCC                                                      & 0.127                                                                         & \textless 0.05   \\ \hline
Avg. NL                                                        & 0.121                                                                         & \textless 0.05   \\ \hline
\rowcolor[HTML]{EFEFEF} 
Avg. NLE                                                       & 0.116                                                                         & \textless 0.05   \\ \hline
Avg. NUMPAR                                                    & -0.037                                                                        & \textless 0.05   \\ \hline
\rowcolor[HTML]{EFEFEF} 
Avg. NOS                                                       & 0.144                                                                         & \textless 0.05   \\ \hline
Avg. NOI                                                       & 0.145                                                                         & \textless 0.05   \\ \hline
\end{tabular}
\end{table}
\begin{table}[ht]
\centering
\caption{Comparing the correlation values of complexity and existence of vulnerability between smart contracts (SC) and traditional programs (TP)}
\label{tab:spearmancompare}
\begin{tabular}{|l|c|c|}
\hline
\rowcolor[HTML]{9B9B9B} 
\multicolumn{1}{|c|}{\cellcolor[HTML]{9B9B9B}\textbf{Metrics}} & \multicolumn{1}{l|}{\cellcolor[HTML]{9B9B9B}\textbf{SC Correlation}} & \textbf{TP Correlation} \\ \hline
SLOC                                                           & 0.153                                                                & \textbf{0.292}          \\ \hline
\rowcolor[HTML]{EFEFEF} 
LLOC                                                           & 0.132                                                                & \textbf{0.288}          \\ \hline
CLOC                                                           & -0.077                                                               & \textbf{0.324}          \\ \hline
\rowcolor[HTML]{EFEFEF} 
NF                                                             & 0.123                                                                & \textbf{0.434}          \\ \hline
WMC                                                            & 0.130                                                                & \textbf{0.429}          \\ \hline
\rowcolor[HTML]{EFEFEF} 
NL                                                             & 0.146                                                                & \textbf{0.532}          \\ \hline
NLE                                                            & 0.144                                                                & -                       \\ \hline
\rowcolor[HTML]{EFEFEF} 
NUMPAR                                                         & 0.082                                                                & -                       \\ \hline
NOS                                                            & 0.160                                                                & \textbf{0.292}          \\ \hline
\rowcolor[HTML]{EFEFEF} 
DIT                                                            & 0.099                                                                & \textbf{0.459}          \\ \hline
NOA                                                            & 0.102                                                                & -                       \\ \hline
\rowcolor[HTML]{EFEFEF} 
NOD                                                            & -0.042                                                               & \textbf{0.642}          \\ \hline
CBO                                                            & 0.182                                                                & \textbf{0.454}          \\ \hline
\rowcolor[HTML]{EFEFEF} 
NA                                                             & 0.170                                                                & -                       \\ \hline
NOI                                                            & 0.165                                                                & \textbf{0.514}          \\ \hline
\rowcolor[HTML]{EFEFEF} 
Avg. McCC                                                      & 0.127                                                                & -                       \\ \hline
Avg. NL                                                        & 0.121                                                                & -                       \\ \hline
\rowcolor[HTML]{EFEFEF} 
Avg. NLE                                                       & 0.116                                                                & -                       \\ \hline
Avg. NUMPAR                                                    & -0.037                                                               & -                       \\ \hline
\rowcolor[HTML]{EFEFEF} 
Avg. NOS                                                       & 0.144                                                                & -                       \\ \hline
Avg. NOI                                                       & 0.145                                                                & -                       \\ \hline
\end{tabular}
\end{table}
\begin{table*}[ht]
\centering
\caption{Statistical comparison (Mann-Whitney U) and Effect Size (Cliff's Delta) between vulnerable and neutral codes}
\label{tab:mannwhit}
\begin{tabular}{|l|c|c|c|c|}
\hline
\rowcolor[HTML]{9B9B9B} 
\multicolumn{1}{|c|}{\cellcolor[HTML]{9B9B9B}\textbf{Metrics}} & \textbf{Discriminative} & \textbf{P-Value}        & \textbf{Cliff's Delta} & \textbf{Effect Size} \\ \hline
SLOC                                                           & yes                     & \textless 0.05 & 0.7069                 & Large                \\ \hline
\rowcolor[HTML]{EFEFEF} 
LLOC                                                           & yes                     & \textless 0.05 & 0.6137                 & Large                \\ \hline
CLOC                                                           & yes                     & \textless 0.05 & -0.2827                & Small                \\ \hline
\rowcolor[HTML]{EFEFEF} 
NF                                                             & yes                     & \textless 0.05 & 0.5676                 & Large                \\ \hline
WMC                                                            & yes                     & \textless 0.05 & 0.5979                 & Large                \\ \hline
\rowcolor[HTML]{EFEFEF} 
NL                                                             & yes                     & \textless 0.05 & 0.5809                 & Large                \\ \hline
NLE                                                            & yes                     & \textless 0.05 & 0.5699                 & Large                \\ \hline
\rowcolor[HTML]{EFEFEF} 
NUMPAR                                                         & yes                     & \textless 0.05 & 0.3782                 & Medium               \\ \hline
NOS                                                            & yes                     & \textless 0.05 & 0.7253                 & Large                \\ \hline
\rowcolor[HTML]{EFEFEF} 
DIT                                                            & yes                     & \textless 0.05 & 0.4118                 & Medium               \\ \hline
NOA                                                            & yes                     & \textless 0.05 & 0.4254                 & Medium               \\ \hline
\rowcolor[HTML]{EFEFEF} 
NOD                                                            & yes                     & \textless 0.05 & -0.1807                & Small                \\ \hline
CBO                                                            & yes                     & \textless 0.05 & 0.5432                 & Large                \\ \hline
\rowcolor[HTML]{EFEFEF} 
NA                                                             & yes                     & \textless 0.05 & 0.7391                 & Large                \\ \hline
NOI                                                            & yes                     & \textless 0.05 & 0.7418                 & Large                \\ \hline
\rowcolor[HTML]{EFEFEF} 
Avg. McCC                                                      & yes                     & \textless 0.05 & 0.5324                 & Large                \\ \hline
Avg. NL                                                        & yes                     & \textless 0.05 & 0.4833                 & Large                \\ \hline
\rowcolor[HTML]{EFEFEF} 
Avg. NLE                                                       & yes                     & \textless 0.05 & 0.4616                 & Medium               \\ \hline
Avg. NUMPAR                                                    & yes                     & \textless 0.05 & -0.1726                & Small                \\ \hline
\rowcolor[HTML]{EFEFEF} 
Avg. NOS                                                       & yes                     & \textless 0.05 & 0.6541                 & Large                \\ \hline
Avg. NOI                                                       & yes                     & \textless 0.05 & 0.6492                 & Large                \\ \hline
\end{tabular}
\end{table*}

\section{Experimental Results}
\begin{figure*}[ht]
\centering
\includegraphics[width=1.0\textwidth]{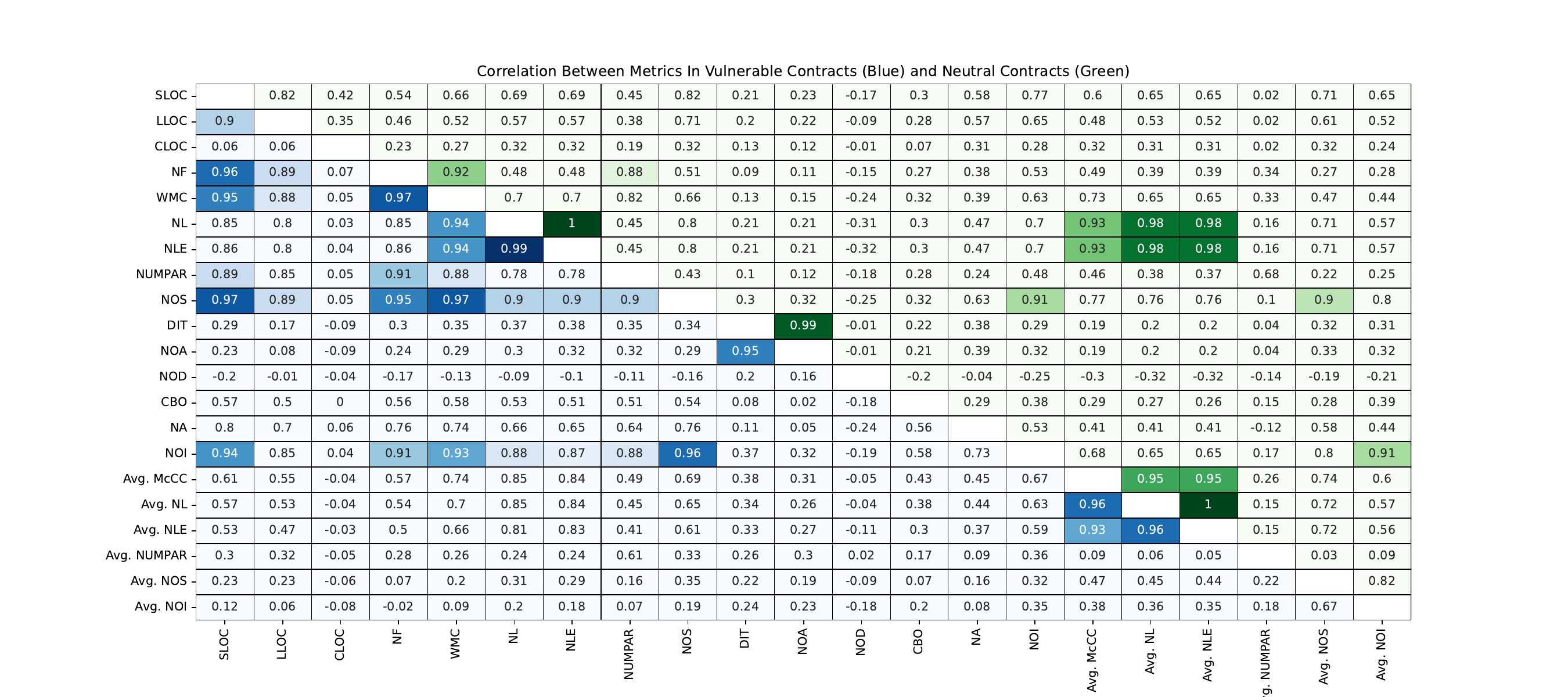}
\caption{Correlation between metrics in vulnerable contracts (blue) and neutral contracts (green)}
\label{fig:q1}
\end{figure*}

In this results section, we give answers to each research question asked. For a full report, you can look at our implementation's source code which is publicly available at  GitHub\footnote{https://github.com/MasoudJTehrani/SC-Complexity-Vulnerability}
\\
\textbf{RQ1: Are there correlations among complexity metrics?}
Our objective was to identify potential redundancies or inconsistencies among the various complexity metrics available. To address this question, we employed Spearman's correlation coefficient. Initially, we divided the codes into two groups: vulnerable codes and neutral codes (without vulnerabilities). Subsequently, we calculated Spearman's correlation coefficient for each metric within each group. This approach enabled us to identify any potential redundancies among the metrics in both code groups. For our analysis, we considered two variables to be strongly correlated if the coefficient ($\rho$) exceeded 0.9~\cite{hinkle2003applied}. Figure~\ref{fig:q1} provides a visual representation of the correlation analysis, distinguishing between vulnerable (blue) and neutral (green) codes. The results of this analysis give information about the interrelationships among the complexity metrics within each code group, helping us understand which metrics are closely related to each other. The highlighted squares in the correlation matrix demonstrate an unusually strong association between certain metrics, suggesting potential redundancy among them.

Among all the coefficients calculated for each group, the p-values for the neutral group were consistently below 0.05, indicating a high level of confidence in the results. However, exceptions were noted in the relationships between CLOC and NOD, DIT and NOD, and NOA and NOD, with p-values of 0.50, 0.09, and 0.09, respectively. Unfortunately, for the vulnerable group, the p-values were higher than 0.05 in many instances. Given the high number of cases, we only report those that are important for our use, specifically when a high correlation is observed.

The potential redundant metrics identified in the analysis are as follows:

\begin{itemize}
    \item \textbf{SLOC:} Strongly related to NOI, NOS, WMC, NF, and LLOC. SLOC counts the number of source lines of code, including comments and empty spaces.
    \item \textbf{NF and WMC:} Conceptually unrelated, but show a relationship. NF counts the number of functions, and WMC calculates McCabe's cyclomatic complexity over functions in a contract. The similarity could be due to the relatively lower number of control structures (to lower gas prices) in smart contracts, resulting in equal numbers for functions and McCabe's complexity. Furthermore, both show a high correlation with similar metrics such as NOI, NOS, NUMPAR, NLE, and NL.
    \item \textbf{NLE, NL, Avg. NLE, and Avg. NL:} These metrics measure the nesting level of control structures. They are equal due to the limited use of ``if-else'' and ``loop'' structures in smart contracts, suggesting that only one of these metrics may be useful.
    \item \textbf{Avg. McCC, Avg. NLE, and Avg. NL:} Similarity due to the same reasons as mentioned before, indicating potential redundancy.
    \item \textbf{NOA and DIT:} These metrics, derived from the inheritance tree, exhibit similarity, possibly indicating redundant information.
    \item \textbf{NOI and Avg. NOI:} These are equal due to the limited use of invocations in smart contracts, suggesting that only one metric might be necessary. The high correlation between NOI and other metrics of SLOC, NF, WMC, and NOS was mentioned previously.
\end{itemize}

\begin{figure}[h!]
\centering
\includegraphics[width=8cm]{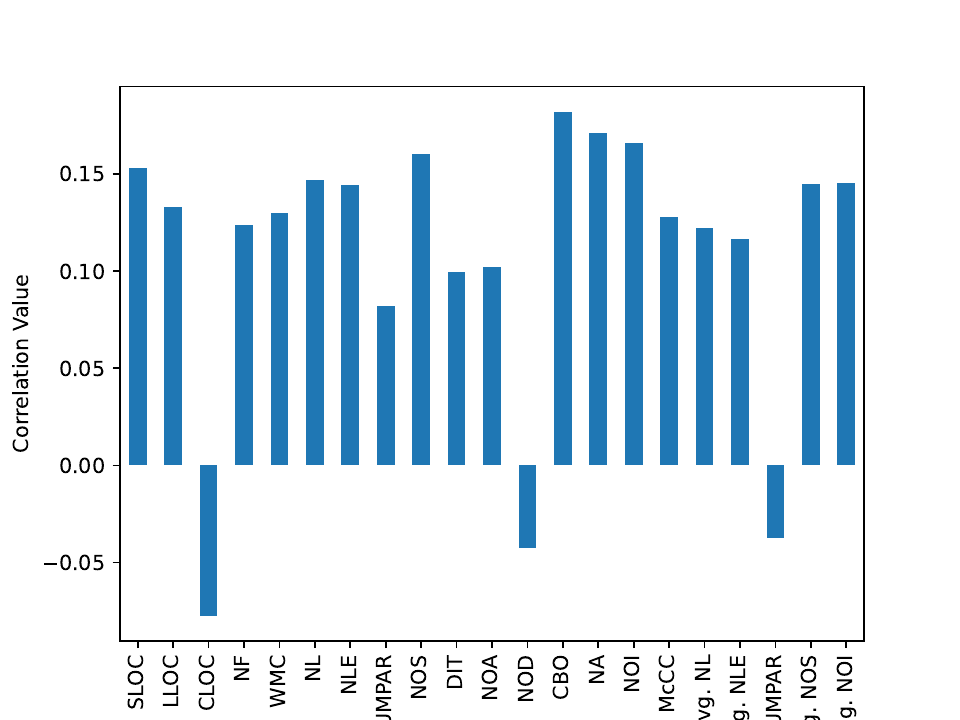}
\caption{Bar plot of Spearman correlations between complexity metrics and vulnerabilities}
\label{fig:spearman}
\end{figure}

Fortunately, all the p-values between these metrics with high correlation were less than 0.05, thus this report is reliable. Considering these potential redundancies, it is crucial to carefully select and include only the most relevant and distinct complexity metrics in the vulnerability prediction models. Removing redundant metrics can optimize the feature set and improve the efficiency and interpretability of the models while achieving comparable predictive performance.
\\
\textbf{RQ2: Is there a correlation between each complexity metric and the existence of vulnerability?}
We aimed to determine whether certain complexity metrics are associated with the presence of vulnerabilities in individual contracts. This part also utilizes the Spearman Correlation Coefficient but without separately grouping the codes into vulnerable and neutral. In our study, all of the calculated p-values for the correlations between complexity metrics and vulnerability indicators were less than 0.05. The result of the Spearman correlation coefficient for each metric is represented in Table~\ref{tab:spearman}. To analyze the result more comprehensively, figure~\ref{fig:spearman} illustrates the result in a bar plot.

The interpretation of the correlation coefficient depends on the context of its usage. According to Cohen et al.,~\cite{cohen2013statistical}, the following guidelines can be used to interpret the strength of the relationship:
\begin{itemize}
    \item A correlation coefficient of less than 0.3 indicates a weak correlation.
    \item A correlation coefficient between 0.3 and 0.5 suggests a medium correlation.
    \item A correlation coefficient greater than 0.5 indicates a strong correlation.
\end{itemize}

These guidelines provide a qualitative assessment of the strength of the relationship between the variables being studied. In our analysis, we apply these interpretations to understand the degree of association between complexity metrics and vulnerability in Solidity smart contracts.
Our analysis reveals that among the complexity metrics used in this study, all metrics show weak correlations with vulnerability in Solidity smart contracts. This finding suggests that no single metric alone can effectively distinguish between vulnerable and neutral codes in smart contracts, which contrasts with previous research on traditional programming languages~\cite{alves2016software, chowdhury2010can} showing stronger correlations between certain complexity metrics and vulnerability, often with coefficients exceeding 0.5.
\\
Table~\ref{tab:spearmancompare} illustrates the comparison of correlation outcomes between complexity metrics in smart contracts and the same metrics in traditional programming languages. Only the metrics consistent with previous research are included. The table findings indicate that while in other languages these metrics exhibited a monotonic relation with vulnerability presence, this relationship was not consistently observed in the context of smart contracts.

\textbf{RQ3: Are the metrics different in vulnerable and neutral
codes?} We sought to investigate whether the complexity metrics can effectively distinguish between functions with reported vulnerabilities and those without reported vulnerabilities. To assess whether the metrics in each group (vulnerable and neutral) can effectively discriminate between vulnerable and neutral codes, we employed the Mann-Whitney U test present in Table~\ref{tab:mannwhit}. This statistical test was conducted by comparing the metric values of the vulnerable group with their corresponding values in the neutral group. \new{While the Mann-Whitney U test indicated statistically significant differences for all metrics ($p < 0.05$), confirming their ability to distinguish between vulnerable and neutral codes, the effect size analysis reveals the strength of these distinctions. Notably, no metric exhibited a negligible effect size, and the majority of complexity metrics demonstrated a large effect size. This reinforces the findings of RQ3, suggesting that most complexity measures are not just statistically significant but are practically robust indicators for identifying vulnerable contracts. Metrics with medium effect sizes, such as DIT ($\delta = 0.4118$), NOA ($\delta = 0.4254$), and Avg. NLE ($\delta = 0.4616$), further support the hypothesis that inheritance depth and nesting levels are consistent discriminators, even if slightly less dominant than the primary complexity indicators.}

\textbf{RQ4: How different are the metrics in vulnerable codes
compared to neutral ones?} We intended to examine whether higher complexity in code is linked to a higher likelihood of vulnerability occurrence. This approach enables a more targeted and efficient allocation of resources toward securing smart contracts.

\begin{figure*}[ht]
\centering
\includegraphics[width=1.0\textwidth]{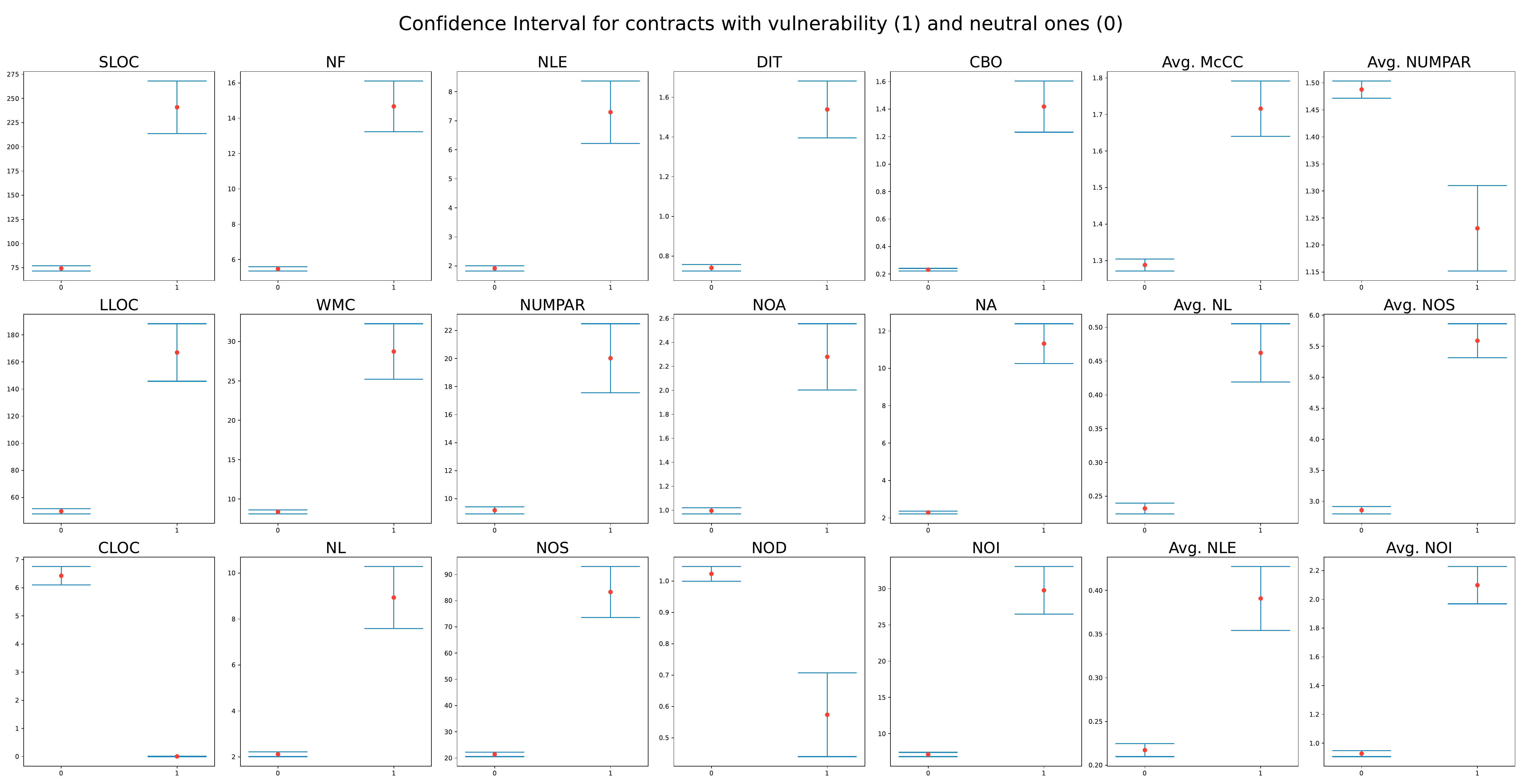}
\caption{Confidence interval comparison in vulnerable contracts (1) and neutral contracts (0)}
\label{fig:q4}
\end{figure*}

\new{To answer this question, we examine the central tendency and spread of each group.} To compare the means of each group, we employed confidence intervals, ensuring that the resulting intervals encompass the true population parameter in around 95\% of cases. This statistical approach allowed us to effectively evaluate the differences between vulnerable and neutral codes in Solidity smart contracts. The comparison results are presented in Figure~\ref{fig:q4}.

In the figure, it can be observed that, apart from the three exceptions, every complexity metric is higher in vulnerable contracts, indicating a positive correlation between complexity and vulnerability. However, the three metrics Avg. NUMPAR, CLOC, and NOD contrast this trend. These findings are distinct from those reported in~\cite{alves2016software}, demonstrating the unique characteristics and complexities present in Solidity smart contracts, especially in the case of CLOC, which measures the number of lines of comments in the code. In previous work, this metric showed higher values in vulnerable codes. However, in our study, higher CLOC values were observed in neutral codes. This suggests that writing comments in smart contracts might help identify and mitigate vulnerabilities within the code.

\new{This finding is also supported by Table~\ref{tab:mannwhit}, with the effect sizes that highlight distinct behaviors among specific features. While most metrics show a strong positive association with vulnerability, CLOC ($\delta = -0.2827$), NOD ($\delta = -0.1807$), and Avg. NUMPAR ($\delta = -0.1726$), exhibited small effect sizes with negative Cliff’s Delta values. This confirms the "inverse" relationship noted in our conclusions: unlike general complexity, vulnerable contracts tend to have fewer comment lines (CLOC) and fewer descendants (NOD) than their neutral counterparts. This nuance is critical for building predictive models, as it suggests that while high complexity generally signals risk, a lack of documentation (comments) or extensibility (descendants) is also a specific, albeit weaker, marker of vulnerability.}
\\
\section{Discussion}
Based on the findings of this study, we can present how complexity metrics can be integrated into existing vulnerability detection tools. Also, we discuss how this study's findings might influence smart contract best practices and developer guidelines. Lastly, we talk about the future directions, and how to extend the study of complexity in smart contracts.

\subsection{Integrating Complexity Metrics}
Complexity metrics can be integrated into existing vulnerability detection tools by leveraging their ability to distinguish between vulnerable and neutral smart contract codes. This study’s findings suggest several practical approaches for enhancing vulnerability detection tools:

\begin{itemize}
    \item \textbf{Feature Optimization:} This study identifies redundancy among complexity metrics, such as strong correlations between SLOC, WMC, and NF. By selecting only the most relevant metrics, vulnerability detection models can be made more efficient, reducing computational overhead without sacrificing accuracy.
    \item \textbf{Risk Assessment Models:} The correlation between certain complexity metrics and vulnerability presence, even if weak, suggests that these metrics can serve as supplementary indicators in risk assessment models. Metrics such as CBO (Coupling Between Objects) and NOI (Number of Invocations) show relatively higher correlations with vulnerabilities and can be weighted accordingly in predictive models.
    \item \textbf{Automated Analyzers:} Smart contract security tools can incorporate discriminative complexity metrics identified in the study (e.g., SLOC, WMC, NOI) as warning signals, flagging contracts with high complexity as more likely to contain vulnerabilities. This aligns with best practices for keeping smart contracts simple and auditable.
    \item \textbf{Static Analysis Enhancements:} Existing static analysis tools can integrate complexity-based heuristics to prioritize contract segments that require deeper analysis. Since vulnerability likelihood increases with complexity, tools can focus audits on high-complexity regions to enhance security assessments.
\end{itemize}

\subsection{Best Practices}
Regarding best practices and developer guidelines, the study’s findings highlight the following key takeaways:

\begin{itemize}
    \item \textbf{Simplicity is Key:} Higher complexity is generally associated with a greater likelihood of vulnerabilities. Developers should aim to minimize unnecessary functions, control structures, and deep nesting to reduce the risk of security flaws.
    \item \textbf{Code Documentation Matters:} Unlike traditional software, smart contracts with higher comment density (CLOC) were found to be less vulnerable, suggesting that well-documented contracts may result in better observation and have fewer security issues.
    \item \textbf{Selective Use of Inheritance:} Metrics related to inheritance (e.g., NOA, DIT) were found to be discriminative, meaning that complex inheritance structures might introduce subtle vulnerabilities. Developers should keep inheritance hierarchies shallow and transparent.
    \item \textbf{Tool Support for Complexity Awareness:} Smart contract development tools and IDEs should integrate complexity visualization and alerts to help developers write safer code by identifying potentially risky patterns early in development.
\end{itemize}

By utilizing these takeaways in vulnerability detection tools and developer guidelines, the security of smart contracts can be significantly improved, leading to more reliable and robust decentralized applications.

\subsection{Future Work}
In future work, we have identified several areas for improvement and expansion in our study. One of the objectives is to include additional complexity metrics that are specific to smart contracts, such as the amount of gas expended, to enhance our analysis.

Another potential expansion is to analyze these metrics for each type of vulnerability separately. This would provide deeper insights into their impact on smart contracts and help developers address vulnerabilities more precisely.

Furthermore, it's reasonable to explore the inclusion of comprehension complexity metrics, such as cognitive complexity~\cite{cognitive2018,campbell2018cognitive}, in our evaluation. Cognitive complexity measures the difficulty of understanding code by considering factors such as nested conditions, control flow, and human cognition. By including such metrics, we can gain a better understanding of the cognitive load imposed on developers and its application to vulnerabilities of Smart Contracts. These future efforts will contribute to a more comprehensive and robust assessment of Smart Contract security.

\new{
Future work should also categorize smart contracts by type (e.g., DeFi, Games, Wallets). A contract with high SLOC in a simple domain might be a better predictor of vulnerability than high SLOC in a complex domain.

To move beyond correlation and establish causality, we propose employing longitudinal analyses of code evolution (tracking metrics as vulnerabilities are introduced or fixed) or quasi-experimental designs where complexity is controlled while monitoring security outcomes.}
\section{Threat To Validity}
In this section, we talk about the potential biases of this paper’s results, how they can affect the results, and how we mitigated them.
\subsection{External Threats}
Given the critical role of datasets in this type of research, any changes in the dataset or labeling could lead to minor variations in experimental outcomes. To mitigate this, this paper utilizes a widely recognized and frequently used dataset in smart contract research. As a result, while some variation may occur with a different dataset, the overall analysis remains valuable and relevant.

The smart contract codes in the dataset were collected using a standard approach from a public resource. Due to the large number of contracts, bias toward a specific Solidity version was minimized. However, since the syntax of older Solidity codes is largely different, collecting older smart contracts and replicating this study on older versions of Solidity may lead to different results.

Solmet was introduced in 2018, but our dataset, collected in 2023, consists primarily of newer smart contracts. This raises concerns about the validity of the complexity metrics measured by Solmet. However, these metrics are not specific to the Solidity programming language and are commonly used in traditional languages, making them unaffected by changes in newer Solidity versions. For instance, metrics related to "for" loops remain consistent across versions, as "for" loops exist in all contract versions. These traditional metrics are used to enable comparisons with traditional programs. The incorporation of smart contract-specific metrics is left for future work.
\subsection{Internal Threats}
This paper employed statistical hypothesis tests, including Mann-Whitney U, to identify significant differences between \new{independent} data sets. Since these tests rely on several assumptions, their reliability, particularly for RQ3, may be affected if those assumptions are not met in a different setup. To mitigate this, these methods were selected based on prior research and the characteristics of the dataset. This also ensured consistency with studies used for comparison.

Similar to the selection of statistical tests, the thresholds and intervals were chosen based on best practices and the dataset. It is reasonable to expect that using different values for these variables could lead to varying results.

\new{We assessed code complexity metrics (e.g., SLOC, WMC) but did not normalize these against the inherent difficulty of the task the smart contract solves. For example, a complex DeFi protocol naturally requires more lines of code and coupling than a simple token contract. Consequently, higher complexity metrics might simply reflect a larger problem domain, which statistically increases the surface area for bugs, rather than indicating that the coding style itself was poor. While we demonstrate a strong association between these metrics and vulnerabilities, future studies should stratify contracts by domain (e.g., DeFi, NFT, Gaming) or normalize metrics against gas consumption profiles to isolate unnecessary implementation complexity from inherent problem complexity.}
\section{Conclusion}
Our study examined the use of complexity metrics as indicators of vulnerable code locations in Solidity smart contracts. We found that vulnerable functions exhibit distinct characteristics in terms of code complexity compared to non-vulnerable functions.\\
Throughout our investigation, we have obtained a valuable understanding of the relationship between complexity metrics and vulnerability detection in Solidity smart contracts. 

\new{It is important to note that these findings represent correlations. High complexity often accompanies larger `problem complexity,' which may be the underlying driver for increased vulnerability risk, rather than the code complexity metrics themselves acting as a causal mechanism.} From our findings, we can draw the following conclusions:

\begin{enumerate}
    \item Some complexity metrics are highly correlated and redundant, suggesting that certain metrics may not provide additional information beyond what others already capture.
    \item Individual complexity metrics alone are insufficient to \new{help} predict vulnerable contracts effectively, as their correlations with the existence of vulnerabilities are weak.
    \item Complexity metrics exhibit distinct values and distributions in vulnerable and non-vulnerable codes, demonstrating their discriminative power.
    \item Most complexity metrics tend to have higher values in vulnerable codes, except for a few specific ones, such as the number of comment lines, number of descendants, and average number of parameters, which were lower in vulnerable codes.
\end{enumerate}

Although the Spearman rank correlation coefficients indicated low individual correlations between each complexity metric and vulnerability, our machine learning models still achieved good results. This suggests that no single metric can effectively distinguish between vulnerable and invulnerable contracts on its own.

This study provides practical evidence of the effectiveness and relevance of complexity metrics in predicting vulnerabilities in Solidity smart contracts and analyzes the power of each metric to help researchers understand their individual influence. Furthermore, this study discusses the methods of how complexity metrics can be integrated into existing vulnerability tools, as well as how developers can influence these findings and implement safer code.

\section*{Data Availability}
The implementations, source code, data, and experimental results are publicly
available at \url{https://github.com/MasoudJTehrani/SC-Complexity-Vulnerability}

\bibliographystyle{IEEEtran}
\bibliography{Bibliography.bib}

@article{buterin2014next,
  title={A next-generation smart contract and decentralized application platform},
  author={Buterin, Vitalik and others},
  journal={white paper},
  volume={3},
  number={37},
  year={2014}
}

@article{szabo1996smart,
  title={Smart contracts: building blocks for digital markets},
  author={Szabo, Nick},
  journal={EXTROPY: The Journal of Transhumanist Thought,(16)},
  volume={18},
  number={2},
  pages={28},
  year={1996}
}

@inproceedings{bartoletti2017empirical,
  title={An empirical analysis of smart contracts: platforms, applications, and design patterns},
  author={Bartoletti, Massimo and Pompianu, Livio},
  booktitle={International conference on financial cryptography and data security},
  pages={494--509},
  year={2017},
  organization={Springer}
}

@article{zhang2017distributed,
  title={Distributed electrical energy systems: Needs, concepts, approaches and vision},
  author={Zhang, Yingchen and Zhang, Jun and Gao, Wenzhong and Zheng, Xinhu and Yang, Liuqing and Hao, Jun and Dai, Xiaoxiao},
  journal={Acta Automatica Sinica},
  volume={43},
  number={NREL/JA-5D00-70646},
  year={2017},
  publisher={National Renewable Energy Lab.(NREL), Golden, CO (United States)}
}

@article{christidis2016blockchains,
  title={Blockchains and smart contracts for the internet of things},
  author={Christidis, Konstantinos and Devetsikiotis, Michael},
  journal={Ieee Access},
  volume={4},
  pages={2292--2303},
  year={2016},
  publisher={Ieee}
}

@article{griggs2018healthcare,
  title={Healthcare blockchain system using smart contracts for secure automated remote patient monitoring},
  author={Griggs, Kristen N and Ossipova, Olya and Kohlios, Christopher P and Baccarini, Alessandro N and Howson, Emily A and Hayajneh, Thaier},
  journal={Journal of medical systems},
  volume={42},
  number={7},
  pages={1--7},
  year={2018},
  publisher={Springer}
}

@inproceedings{alhadhrami2017introducing,
  title={Introducing blockchains for healthcare},
  author={Alhadhrami, Zainab and Alghfeli, Salma and Alghfeli, Mariam and Abedlla, Juhar Ahmed and Shuaib, Khaled},
  booktitle={2017 international conference on electrical and computing technologies and applications (ICECTA)},
  pages={1--4},
  year={2017},
  organization={IEEE}
}

@article{jeon2022blockchain,
  title={Blockchain and AI Meet in the Metaverse},
  author={Jeon, Hyun-joo and Youn, Ho-chang and Ko, Sang-mi and Kim, Tae-heon},
  journal={Advances in the Convergence of Blockchain and Artificial Intelligence},
  pages={73},
  year={2022},
  publisher={BoD--Books on Demand}
}

@inproceedings{bui2021evaluating,
  title={Evaluating Upgradable Smart Contract},
  author={Bui, Van Cuong and Wen, Sheng and Yu, Jiangshan and Xia, Xin and Haghighi, Mohammad Sayad and Xiang, Yang},
  booktitle={2021 IEEE International Conference on Blockchain (Blockchain)},
  pages={252--256},
  year={2021},
  organization={IEEE}
}

@article{wang2021ethereum,
  title={Ethereum smart contract security research: survey and future research opportunities},
  author={Wang, Zeli and Jin, Hai and Dai, Weiqi and Choo, Kim-Kwang Raymond and Zou, Deqing},
  journal={Frontiers of Computer Science},
  volume={15},
  number={2},
  pages={1--18},
  year={2021},
  publisher={Springer}
}

@inproceedings{atzei2017survey,
  title={A survey of attacks on ethereum smart contracts (sok)},
  author={Atzei, Nicola and Bartoletti, Massimo and Cimoli, Tiziana},
  booktitle={International conference on principles of security and trust},
  pages={164--186},
  year={2017},
  organization={Springer}
}

@article{vivar2020analysis,
  title={An analysis of smart contracts security threats alongside existing solutions},
  author={Vivar, Antonio Lopez and Castedo, Alberto Tur{\'e}gano and Orozco, Ana Lucila Sandoval and Villalba, Luis Javier Garc{\'\i}a},
  journal={Entropy},
  volume={22},
  number={2},
  year={2020},
  publisher={Multidisciplinary Digital Publishing Institute (MDPI)}
}

@inproceedings{shin2008complexity,
  title={Is complexity really the enemy of software security?},
  author={Shin, Yonghee and Williams, Laurie},
  booktitle={Proceedings of the 4th ACM workshop on Quality of protection},
  pages={47--50},
  year={2008}
}

@article{hegedHus2019towards,
  title={Towards analyzing the complexity landscape of solidity based ethereum smart contracts},
  author={Heged{\H{u}}s, P{\'e}ter},
  journal={Technologies},
  volume={7},
  number={1},
  pages={6},
  year={2019},
  publisher={Multidisciplinary Digital Publishing Institute}
}

@inproceedings{feist2019slither,
  title={Slither: a static analysis framework for smart contracts},
  author={Feist, Josselin and Grieco, Gustavo and Groce, Alex},
  booktitle={2019 IEEE/ACM 2nd International Workshop on Emerging Trends in Software Engineering for Blockchain (WETSEB)},
  pages={8--15},
  year={2019},
  organization={IEEE}
}

@article{harer2018automated,
  title={Automated software vulnerability detection with machine learning},
  author={Harer, Jacob A and Kim, Louis Y and Russell, Rebecca L and Ozdemir, Onur and Kosta, Leonard R and Rangamani, Akshay and Hamilton, Lei H and Centeno, Gabriel I and Key, Jonathan R and Ellingwood, Paul M and others},
  journal={arXiv preprint arXiv:1803.04497},
  year={2018}
}

@article{mehar2019understanding,
  title={Understanding a revolutionary and flawed grand experiment in blockchain: the DAO attack},
  author={Mehar, Muhammad Izhar and Shier, Charles Louis and Giambattista, Alana and Gong, Elgar and Fletcher, Gabrielle and Sanayhie, Ryan and Kim, Henry M and Laskowski, Marek},
  journal={Journal of Cases on Information Technology (JCIT)},
  volume={21},
  number={1},
  pages={19--32},
  year={2019},
  publisher={IGI Global}
}

@inproceedings{shin2011initial,
  title={An initial study on the use of execution complexity metrics as indicators of software vulnerabilities},
  author={Shin, Yonghee and Williams, Laurie},
  booktitle={Proceedings of the 7th International workshop on software engineering for secure systems},
  pages={1--7},
  year={2011}
}

@article{moshtari2013using,
  title={Using complexity metrics to improve software security},
  author={Moshtari, Sara and Sami, Ashkan and Azimi, Mahdi},
  journal={Computer Fraud \& Security},
  volume={2013},
  number={5},
  pages={8--17},
  year={2013},
  publisher={Elsevier}
}

@inproceedings{akca2019solanalyser,
  title={SolAnalyser: A framework for analysing and testing smart contracts},
  author={Akca, Sefa and Rajan, Ajitha and Peng, Chao},
  booktitle={2019 26th Asia-Pacific Software Engineering Conference (APSEC)},
  pages={482--489},
  year={2019},
  organization={IEEE}
}

@article{grishchenko2018ethertrust,
  title={Ethertrust: Sound static analysis of ethereum bytecode},
  author={Grishchenko, Ilya and Maffei, Matteo and Schneidewind, Clara},
  journal={Technische Universit{\"a}t Wien, Tech. Rep},
  pages={1--41},
  year={2018}
}

@article{mccabe1976complexity,
  title={A complexity measure},
  author={McCabe, Thomas J},
  journal={IEEE Transactions on software Engineering},
  number={4},
  pages={308--320},
  year={1976},
  publisher={IEEE}
}

@inproceedings{chowdhury2010can,
  title={Can complexity, coupling, and cohesion metrics be used as early indicators of vulnerabilities?},
  author={Chowdhury, Istehad and Zulkernine, Mohammad},
  booktitle={Proceedings of the 2010 ACM Symposium on Applied Computing},
  pages={1963--1969},
  year={2010}
}

@article{chowdhury2011using,
  title={Using complexity, coupling, and cohesion metrics as early indicators of vulnerabilities},
  author={Chowdhury, Istehad and Zulkernine, Mohammad},
  journal={Journal of Systems Architecture},
  volume={57},
  number={3},
  pages={294--313},
  year={2011},
  publisher={Elsevier}
}

@inproceedings{almogahed2022software,
  title={Software Security Measurements: A Survey},
  author={Almogahed, Abdullah and Omar, Mazni and Zakaria, Nur Haryani and Alawadhi, Abdulwadood},
  booktitle={2022 International Conference on Intelligent Technology, System and Service for Internet of Everything (ITSS-IoE)},
  pages={1--6},
  year={2022},
  organization={IEEE}
}

@article{mcgraw2006software,
  title={Software security},
  author={McGraw, Gary},
  journal={Building security in},
  year={2006}
}

@book{schneier2003beyond,
  title={Beyond fear: Thinking sensibly about security in an uncertain world},
  author={Schneier, Bruce and Schneier, B},
  volume={10},
  year={2003},
  publisher={Springer}
}

@article{shin2010evaluating,
  title={Evaluating complexity, code churn, and developer activity metrics as indicators of software vulnerabilities},
  author={Shin, Yonghee and Meneely, Andrew and Williams, Laurie and Osborne, Jason A},
  journal={IEEE transactions on software engineering},
  volume={37},
  number={6},
  pages={772--787},
  year={2010},
  publisher={IEEE}
}

@misc{cognitive2018,
  title={Cognitive complexity - a new way of measuring understandability},
  author={Campbell, G Ann},
  year={2018},
  url={https://www.sonarsource.com/docs/CognitiveComplexity.pdf}
}

@inproceedings{campbell2018cognitive,
  title={Cognitive complexity: An overview and evaluation},
  author={Campbell, G Ann},
  booktitle={Proceedings of the 2018 International Conference on Technical Debt},
  pages={57--58},
  year={2018}
}

@inproceedings{peitek2021program,
  title={Program comprehension and code complexity metrics: An fmri study},
  author={Peitek, Norman and Apel, Sven and Parnin, Chris and Brechmann, Andr{\'e} and Siegmund, Janet},
  booktitle={2021 IEEE/ACM 43rd International Conference on Software Engineering (ICSE)},
  pages={524--536},
  year={2021},
  organization={IEEE}
}

@inproceedings{malik2008understanding,
  title={Understanding the rationale for updating a function’s comment},
  author={Malik, Haroon and Chowdhury, Istehad and Tsou, Hsiao-Ming and Jiang, Zhen Ming and Hassan, Ahmed E},
  booktitle={2008 IEEE International Conference on Software Maintenance},
  pages={167--176},
  year={2008},
  organization={IEEE}
}

@article{chidamber1994metrics,
  title={A metrics suite for object oriented design},
  author={Chidamber, Shyam R and Kemerer, Chris F},
  journal={IEEE Transactions on software engineering},
  volume={20},
  number={6},
  pages={476--493},
  year={1994},
  publisher={IEEE}
}

@article{basili1980qualitative,
  title={Qualitative software complexity models: A summary},
  author={Basili, Victor R},
  journal={Tutorial on models and methods for software management and engineering},
  year={1980},
  publisher={IEEE Computer Society Press Los Alamitos, CA}
}

@article{kearney1986software,
  title={Software complexity measurement},
  author={Kearney, Joseph P and Sedlmeyer, Robert L and Thompson, William B and Gray, Michael A and Adler, Michael A},
  journal={Communications of the ACM},
  volume={29},
  number={11},
  pages={1044--1050},
  year={1986},
  publisher={ACM New York, NY, USA}
}

@book{cohen2013statistical,
  title={Statistical power analysis for the behavioral sciences},
  author={Cohen, Jacob},
  year={2013},
  publisher={Academic press}
}

@inproceedings{alves2016software,
  title={Software metrics and security vulnerabilities: dataset and exploratory study},
  author={Alves, Henrique and Fonseca, Baldoino and Antunes, Nuno},
  booktitle={2016 12th European Dependable Computing Conference (EDCC)},
  pages={37--44},
  year={2016},
  organization={IEEE}
}

@article{hinkle2003applied,
  title={Applied statistics for the behavioral sciences},
  author={Hinkle, Dennis E and Wiersma, William and Jurs, Stephen G},
  journal={(No Title)},
  year={2003}
}

@book{moore2007basic,
  title={The basic practice of statistics},
  author={Moore, David S and Kirkland, Stephane},
  volume={2},
  year={2007},
  publisher={WH Freeman New York}
}

@article{liu2023rethinking,
  title={Rethinking Smart Contract Fuzzing: Fuzzing With Invocation Ordering and Important Branch Revisiting},
  author={Liu, Zhenguang and Qian, Peng and Yang, Jiaxu and Liu, Lingfeng and Xu, Xiaojun and He, Qinming and Zhang, Xiaosong},
  journal={arXiv preprint arXiv:2301.03943},
  year={2023}
}

@article{sultana2021using,
  title={Using software metrics for predicting vulnerable classes and methods in Java projects: A machine learning approach},
  author={Sultana, Kazi Zakia and Anu, Vaibhav and Chong, Tai-Yin},
  journal={Journal of Software: Evolution and Process},
  volume={33},
  number={3},
  pages={e2303},
  year={2021},
  publisher={Wiley Online Library}
}

@article{kushwaha2022ethereum,
  title={Ethereum smart contract analysis tools: A systematic review},
  author={Kushwaha, Satpal Singh and Joshi, Sandeep and Singh, Dilbag and Kaur, Manjit and Lee, Heung-No},
  journal={Ieee Access},
  volume={10},
  pages={57037--57062},
  year={2022},
  publisher={IEEE}
}

@inproceedings{agarwal2022cyclomatic,
  title={Cyclomatic complexity analysis for smart contract using control flow graph},
  author={Agarwal, Shantanu and Godboley, Sangharatna and Krishna, P Radha},
  booktitle={International Conference on Computing, Communication and Learning},
  pages={65--78},
  year={2022},
  organization={Springer}
}

@article{tonelli2023smart,
  title={Smart contracts software metrics: A first study},
  author={Tonelli, Roberto and Pierro, Giuseppe Antonio and Ortu, Marco and Destefanis, Giuseppe},
  journal={Plos one},
  volume={18},
  number={4},
  pages={e0281043},
  year={2023},
  publisher={Public Library of Science San Francisco, CA USA}
}

@article{mann1947test,
  title={On a test of whether one of two random variables is stochastically larger than the other},
  author={Mann, Henry B and Whitney, Donald R},
  journal={The annals of mathematical statistics},
  pages={50--60},
  year={1947},
  publisher={JSTOR}
}

@article{romano2006appr,
author = {Romano, Jeanine and Kromrey, Jeffrey},
year = {2006},
month = {01},
pages = {},
title = {Appropriate Statistics for Ordinal Level Data: Should We Really Be Using t-test and Cohen's d for Evaluating Group Differences on the NSSE and other Surveys?}
}

@article{MeisselYao2024CliffsDelta,
  author  = {Meissel, Kane and Yao, Esther S.},
  title   = {Using Cliff's Delta as a Non-Parametric Effect Size Measure: An Accessible Web App and R Tutorial},
  journal = {Practical Assessment, Research, and Evaluation},
  volume  = {29},
  number  = {1},
  year    = {2024}
}

\newpage

\vfill

\end{document}